\documentclass[conference,a4paper]{APSIPA2020}
\usepackage{multirow}
\usepackage{graphicx}
\usepackage{amsmath}
\usepackage[psamsfonts]{amssymb}
\usepackage{amsxtra}
\usepackage{threeparttable}

\usepackage{cite}
\usepackage{bm}
\usepackage{amssymb}
\usepackage{booktabs}

\newcommand{\bs}[1]{\boldsymbol {#1}}
\newcommand{\sigmoid}{f_{\sigma}}

\DeclareMathOperator{\diag}{diag}

\begin{document}

\title{Gamma Boltzmann Machine for Simultaneously Modeling Linear- and Log-amplitude Spectra}

\author{%
\authorblockN{%
Toru Nakashika\authorrefmark{1} and
Kohei Yatabe\authorrefmark{2} \hspace{2pt}
}
\authorblockA{%
\authorrefmark{1}
University of Electro-Communications, Tokyo, Japan \hspace{6pt}}
\authorblockA{%
\authorrefmark{2}
Waseda University, Tokyo, Japan \hspace{6pt}}
}

\maketitle
\thispagestyle{empty}

\begin{abstract}
    In audio applications, one of the most important representations of audio signals is the amplitude spectrogram.
    It is utilized in many machine-learning-based information processing methods including the ones using the restricted Boltzmann machines (RBM).
    However, the ordinary Gaussian-Bernoulli RBM (the most popular RBM among its variations) cannot directly handle amplitude spectra because the Gaussian distribution is a symmetric model allowing negative values which never appear in the amplitude.
    In this paper, after proposing a general gamma Boltzmann machine, we propose a practical model called the \textit{gamma-Bernoulli RBM} that simultaneously handles both linear- and log-amplitude spectrograms.
    Its conditional distribution of the observable data is given by the \textit{gamma distribution}, and thus the proposed RBM can naturally handle the data represented by positive numbers as the amplitude spectra.
    It can also treat amplitude in the logarithmic scale which is important for audio signals from the perceptual point of view.
    The advantage of the proposed model compared to the ordinary Gaussian-Bernoulli RBM was confirmed by PESQ and MSE in the experiment of representing the amplitude spectrograms of speech signals.
\end{abstract}

\section{Introduction}

The Boltzmann machine \cite{ackley1985learning} is a famous stochastic neural network that can discover data representations, in terms of probability distribution, without supervision.
Its variant called the restricted Boltzmann machine (RBM) \cite{Freund:1994tu} is of great practical importance because RBM can be trained with less computational effort than the other models.
Owing to its capability in discovering latent representations without labeled data, RBM has been successfully utilized in various applications involving pattern recognition and machine learning, including computer vision \cite{Lee:2008uz}, collaborative filtering \cite{salakhutdinov2007restricted}, and even geochemical analysis \cite{chen2014application}, to name a few.

In audio applications of RBM, signals are usually modeled based on their amplitude spectra.
Since audio signals can be well characterized by their spectral components, RBM is trained to approximate the probability distribution of the given data in the domain related to frequency.
For example, many studies have applied RBM to model the mel-frequency cepstral coefficients (MFCC) \cite{mohamed2010phone,oo2016dnn} or mel-cepstral features \cite{hu2016dbn,nakashika2016non} of speech signals.
Raw amplitude or STRAIGHT \cite{kawahara2008tandem} spectra have also been considered for extracting richer information from the signals \cite{li2015feature,ling2013modeling,nakashika2018lstbm}.
Moreover, some studies attempted modeling the raw signals using RBM \cite{jaitly2011learning,sailor2016novel}.
All of these are not easy tasks for the original RBM defined for binary signals, and therefore the Gaussian-Bernoulli RBM \cite{hinton2006reducing,cho2011improved} is usually chosen for the audio applications because it can naturally handle real-valued data.

However, modeling of amplitude spectra by the Gaussian-Bernoulli RBM has two issues from the viewpoint of audio applications.
First, the Gaussian distribution allows negative values that are not consistent with the concept of amplitude.
Since amplitude spectra are calculated via absolute value, they are always nonnegative by definition.
Handling nonnegative values with the Gaussian distribution is not straightforward, and therefore the learned representation may contain unavoidable model error.
Second, the human auditory system recognizes sound in the logarithmic-like scale rather than the linear scale.
Based on this fact, many handcrafted audio features as MFCC involves the logarithmic operation within their calculation processes.
Although usefulness of log-amplitude spectra is well-known in the literature, the asymmetric nature of the logarithmic function may make the training difficult for symmetric models as the Gaussian distribution.
Moreover, log-amplitude of approximately sparse spectra (e.g., those of many audio signals including speech) can cause extreme outliers when the amplitude is around zero.
These issues should be resolved for better modeling of audio signals.

In this paper, we propose the \textit{gamma-Bernoulli RBM} for explicitly modeling linear- and log-amplitude spectrograms.
At first, a general gamma Boltzmann machine is defined by a new energy function consisting of the usual quadratic term and an additional log-amplitude term.
Such addition enables simultaneous consideration of the linear- and log-amplitude spectrograms.
Then, its connection is restricted to form the gamma-Bernoulli RBM.
The proposed model represents the conditional distribution of the visible units by the gamma distribution which naturally limits the domain of data to positive numbers.
Owing to these properties, the gamma-Bernoulli RBM should be suitable for representing amplitude spectra and hence audio signals.
By the experiment of reconstructing amplitude spectrograms, the effectiveness of the proposed RBM compared to the ordinary Gaussian-Bernoulli RBM was confirmed in terms of PESQ and mean squared error (MSE).

\section{Boltzmann Machines}
In this section, the ordinary Boltzmann machines are briefly reviewed for contrasting the difference between the conventional and proposed models.

\subsection{Boltzmann Machine}
The Boltzmann machine \cite{ackley1985learning} is an unsupervised neural network for approximating a distribution of the given data.
Let $\bs{x}\in\mathcal{X}$ be a vector, where $\mathcal{X}$ is a space of the variables under investigation (they will be clarified later).
Then, a Boltzmann machine represents its probability density function (PDF) as
\begin{equation}
    p(\bs{x}) = \frac{1}{Z} e^{-E(\bs{x})},
    \label{eq:defPDFx}
\end{equation}
where $E(\,\cdot\,)$ is the so-called energy function, and $Z = \int_\mathcal{X} e^{-E(\bs{x})} \,\mathrm{d}\bs{x}$ is the normalizing constant called partition function.
A type of Boltzmann machines is determined by the definition of the energy function.
In this section, the following energy function involving the parameters $\mathbf{U}$ and $\mathbf{u}$ is considered for explaining the conventional models:
\begin{equation}
    E(\bs{x}) = - \frac{1}{2} \bs{x}^\mathsf{T} \mathbf{U} \bs{x} - \mathbf{u}^\mathsf{T} \bs{x},
    \label{eq:originalBM}
\end{equation}
where the explicit forms of the parameters are given later.

\subsection{Restricted Boltzmann Machine (RBM)}
RBM is the most important variants of the Boltzmann machine.
The above general Boltzmann machine may not be practical because the calculation (or even approximation) of the integral is quite difficult, which makes its training extremely slow for practical dimensionality.
To avoid such difficulty, RBM restricts the connection between the units so that a fast training algorithm can be developed.

In RBM, the variables are separated into two: the visible and hidden variables denoted by $\bs{v}$ and $\bs{h}$, respectively.
An element of these vectors is called a unit, and their connection is defined by the energy function.
$\bs{v}$ corresponds to the data points (and hence \textit{visible}), while $\bs{h}$ represents the latent variables for conditional hidden representation of the data.
That is, a PDF of the visible data is given by the following marginalization:
\begin{equation}
    p(\bs{v}) = \int_\mathcal{H}p(\bs{v},\bs{h})\,\mathrm{d}\bs{h} = \frac{1}{Z}\int_\mathcal{H} e^{-E(\bs{v},\bs{h})}\, \mathrm{d}\bs{h},
    \label{eq:defMarginalPvh}
\end{equation}
where $Z=\int_{\mathcal{V}\times\mathcal{H}}e^{-E(\bs{v},\bs{h})}\,\mathrm{d}\bs{v}\mathrm{d}\bs{h}$, and $\mathcal{V}$ and $\mathcal{H}$ are the spaces of visible and hidden variables, respectively.

The energy function of RBM $E(\bs{v},\bs{h})$ is \textit{restricted} so that both visible and hidden units do not have interconnections (i.e., RBM does not have visible-visible and hidden-hidden connections that can be introduced through the energy function by adding $\bs{v}^\mathsf{T}\mathbf{W}\bs{v}$ and $\bs{h}^\mathsf{T}\mathbf{W}\bs{h}$, respectively, with a square matrix $\mathbf{W}$ having non-diagonal elements).
Such restriction enables fast training by sampling from the conditional distributions: $p(\bs{v}|\bs{h})$ and $p(\bs{h}|\bs{v})$.
These two conditional probabilities are the key ingredients for characterizing the types of RBMs.

\subsection{Bernoulli-Bernoulli RBM}
The original RBM \cite{Freund:1994tu} was defined for binary variables, i.e., $\mathcal{V}$ and $\mathcal{H}$ are the sets of binary numbers: $\bs{v} \in \{0,1\}^I$, $\bs{h} \in \{0,1\}^J$.
The energy function $E_\mathrm{B}(\bs{v},\bs{h})$ is defined as
\begin{equation}
    E_\mathrm{B}(\bs{v},\bs{h}) = - \bs{v}^\mathsf{T} \mathbf{W} \bs{h} - \bs{b}^\mathsf{T} \bs{v} - \bs{c}^\mathsf{T} \bs{h},
    \label{eq:energyBBRBM}
\end{equation}
that is related to the general Boltzmann machine in Eq.~\eqref{eq:originalBM} as
\begin{equation}
    \bs{x} = \left[\begin{array}{c} \bs{v} \\ \bs{h} \end{array}\right], \;\quad\;
    \mathbf{U} = \left[\begin{array}{cc} \!\!\mathbf{O} & \mathbf{W} \\ \mathbf{W}^\mathsf{T}\! & \mathbf{O} \end{array}\right], \;\quad\;
    \bs{u} = \left[\begin{array}{c} \bs{b} \\ \bs{c} \end{array}\right],
\end{equation}
where $\mathbf{W}\in\mathbb{R}^{I\times J}$, $\bs{b}\in\mathbb{R}^I$, $\bs{c}\in\mathbb{R}^J$, $\mathbf{O}$ represents the all-zero matrix with appropriate size, and the operations between the binary and real numbers are performed by regarding the binary symbols as real numbers.

This type of RBM is called the Bernoulli-Bernoulli RBM because the two conditional probabilities required for its training are the element-wise Bernoulli distributions $\mathcal{B}(\:\cdot\:;\bs{p})$:
\begin{align}
    p(\bs{v}|\bs{h}) &= \mathcal{B}(\bs{v}; \sigmoid[\,\bs{b} + \mathbf{W} \bs{h}\,]), \\
    p(\bs{h}|\bs{v}) &= \mathcal{B}(\bs{h}; \sigmoid[\,\bs{c} + \mathbf{W}^\mathsf{T} \bs{v}\,]),
\end{align}
where $\bs{p}\in[0,1]^{I}$ (or $[0,1]^{J}$) is a vector representing the probabilities of taking the value $1$ for each element, and $\sigmoid[\,\cdot\,]$ denotes the element-wise sigmoid function.

\subsection{Gaussian-Bernoulli RBM}
The Bernoulli-Bernoulli RBM has an obvious limitation that the visible variables must be binary.
That is, it can only handle binary data even though many of the interesting real-world data are apparently not binary in nature.
In this respect, the Gaussian-Bernoulli RBM \cite{hinton2006reducing} is the most important variants of RBM because it can naturally handle real-valued data $\bs{v} \in \mathbb{R}^I$, while the hidden variables are remained binary, $\bs{h} \in \{0,1\}^J$.
The energy function $E_\mathrm{G}(\bs{v},\bs{h})$ is defined as%
\footnote{
Note that this definition is somewhat different from those defined in \cite{hinton2006reducing} or \cite{cho2011improved}.
We defined the energy function as in Eq.~\eqref{eq:energyGBRBM} because we empirically found that this works better for our application in Section~\ref{sec:experiments}.
}
\begin{equation}
    E_\mathrm{G}(\bs{v},\bs{h}) = \frac{1}{2} \bs{v}^\mathsf{T} \mathbf{\Sigma}^{-1} \bs{v} - \bs{v}^\mathsf{T} \mathbf{W} \bs{h} - \bs{b}^\mathsf{T} \bs{v} - \bs{c}^\mathsf{T} \bs{h},
    \label{eq:energyGBRBM}
\end{equation}
that is related to the general Boltzmann machine in Eq.~\eqref{eq:originalBM} as
\begin{equation}
    \bs{x} = \left[\begin{array}{c} \bs{v} \\ \bs{h} \end{array}\right], \,\quad
    \mathbf{U} = \left[\begin{array}{cc} \!\!-\mathbf{\Sigma}^{-1}\! & \mathbf{W} \\ \mathbf{W}^\mathsf{T}\! & \mathbf{O} \end{array}\right], \,\quad
    \bs{u} = \left[\begin{array}{c} \bs{b} \\ \bs{c} \end{array}\right],
\end{equation}
where $\mathbf{\Sigma} = \diag(\bs{\sigma}^2)$ is a diagonal matrix, $\bs{\sigma}^2\in\mathbb{R}_{++}^I$ is the model parameter representing variance of the visible variables, and $\diag(\cdot)$ is the operator constructing the diagonal matrix from an input vector.
Its difference from that of the Bernoulli-Bernoulli RBM in Eq.~\eqref{eq:energyBBRBM} is merely the first term $\bs{v}^\mathsf{T} \mathbf{\Sigma}^{-1} \bs{v}$ which represents the self-connection of the visible units.
Note that this term does not introduce interconnection of the visible units because the matrix $\mathbf{\Sigma}^{-1}$ does not have any non-diagonal element.

This model defined by Eq.~\eqref{eq:energyGBRBM} is called the Gaussian-Bernoulli RBM because its conditional probabilities are
\begin{align}
p(\bs{v}|\bs{h}) &= \mathcal{N}(\bs{v}; \mathbf{\Sigma}(\bs{b} + \mathbf{W} \bs{h}), \mathbf{\Sigma}), \\
p(\bs{h}|\bs{v}) &= \mathcal{B}(\bs{h}; \sigmoid[\,\bs{c} + \mathbf{W}^\mathsf{T} \bs{v}\,]),
\end{align}
where $\mathcal{N}(\:\cdot\:; \bs{\mu}, \mathbf{S})$ is the Gaussian distribution with a mean vector $\bs{\mu}\in\mathbb{R}^I$ and a covariance matrix $\mathbf{S}\in\mathbb{R}_{++}^{I\times I}$.
That is, the data are handled by the Gaussian distribution, while the hidden variables are by the Bernoulli distribution.
Therefore, it can approximate the distribution of real-valued data by learning the parameters $(\mathbf{\Sigma},\mathbf{W},\bs{b},\bs{c})$ from the given data.

\section{Gamma Boltzmann Machine}
Among Boltzmann machines, the Gaussian-Bernoulli RBM has been a standard choice for many real-world applications because it can handle real-valued signals.
In audio applications, the amplitude spectrogram is one of the most reasonable choices of a meaningful acoustic feature, and therefore its generative modeling has been investigated \cite{nakashika2018lstbm,vasquez2019melnet,kumar2019melgan,wang2017tacotron,kaneko2017generative}.
However, as mentioned in the Introduction (3rd paragraph), modeling of amplitude spectrograms by the Gaussian-Bernoulli RBM has two issues: production of negative values and ignorance of the human auditory mechanism.
To circumvent these issues, we propose a new variant of the Boltzmann machines named the \textit{gamma-Bernoulli RBM} in this section.

\subsection{Proposed Gamma Boltzmann Machine}
Similarly to the previous section, we first define a general Boltzmann machine without the restriction.
We propose the generative model termed \textit{gamma Boltzmann machine} by defining the following energy function involving logarithmic terms:
\begin{align}
    E_{\log}(\bs{x}) = - \frac{1}{2} &\bs{x}^\mathsf{T} \mathbf{U} \bs{x}  - \bs{u}^\mathsf{T} \bs{x}\nonumber\\
    &- \frac{1}{2} \log(\bs{x})^\mathsf{T} \mathbf{S} \log(\bs{x})  - \bs{s}^\mathsf{T} \log(\bs{x}),
    \label{eq:defGammaBoltzmannMachine}
\end{align}
where $\log(\cdot)$ is the element-wise logarithmic function, $\bs{x}$ is a positive vector (i.e., $x_{n\!}>0$ $\,\forall n\,$), and its PDF is given by Eq.~\eqref{eq:defPDFx}: $p(\bs{x}) =  \exp(-E(\bs{x}))/Z$.
Owing to the existence of $\log(\bs{x})$, this model naturally enforces the variables $\bs{x}$ to be positive.
By introducing the log-related parameters $\mathbf{S}$ and $\bs{s}$, it can learn a PDF with consideration of the logarithmic scale.

\subsection{Proposed Gamma-Bernoulli RBM}
By introducing the visible and hidden units and imposing the restriction, we can obtain RBM based on the above gamma Boltzmann machine.
Due to the logarithmic function in Eq.~\eqref{eq:defGammaBoltzmannMachine}, all variables must be positive.
In our model, the data are assumed to be positive, $\bs{v} \in \mathbb{R}_{++}^I$, and the hidden variables are binary, $\bs{h} \in \{0,1\}^J$.
However, this assumption cannot be accepted directly because $\log(\bs{h})$ takes $-\infty$ whenever $\bs{h}$ contains $0$.
Therefore, we consider transformation that makes the values positive: $\exp(\bs{h})\in\{1,e\}^J$, where $\exp(\cdot)$ for a vector input is the element-wise exponential function.
With this modification, the energy function $E_\Gamma(\bs{v},\bs{h})$ is defined as
\begin{align}
    E_\Gamma(\bs{v},\bs{h}) = &- \bs{v}^\mathsf{T} \mathbf{W} \exp(\bs{h}) - \bs{c}^\mathsf{T} \exp(\bs{h}) \nonumber\\
    &- \log(\bs{v})^\mathsf{T} (\mathbf{V} \bs{h}-\bs{1}) - \bs{d}^\mathsf{T} \bs{h},
    \label{eq:defGammaRBMenergy}
\end{align}
that can be derived from Eq.~\eqref{eq:defGammaBoltzmannMachine} by inserting
\begin{align}
    \bs{x} = \left[\begin{array}{c} \bs{v} \\ \!\!\exp(\bs{h})\!\! \end{array}\right], \quad
    \mathbf{U} &= \left[\!\begin{array}{cc} \!\!\mathbf{O} & \mathbf{W} \\ \mathbf{W}^\mathsf{T}\! & \mathbf{O} \end{array}\!\right], \:
    &\bs{u} &= \left[\begin{array}{c} \bs{0} \\ \bs{c} \end{array}\right], \\
    \mathbf{S} &= \left[\begin{array}{cc} \!\!\mathbf{O} & \mathbf{V} \\ \mathbf{V}^\mathsf{T}\! & \mathbf{O} \end{array}\right], \;\:
    &\bs{s} &= \left[\begin{array}{c} \!\!-\bs{1}\!\! \\ \bs{d} \end{array}\right],\!
    \label{eq:defSs}
\end{align}
where $-\mathbf{W}\in\mathbb{R}^{I\times J}_{++}$, $\mathbf{V}\in\mathbb{R}^{I\times J}_{++}$, $\bs{1}\in\{1\}^I$, $\bs{d}\in\mathbb{R}^J$, $\bs{0}$ represents the all-zero vector with appropriate size, and the joint density function of the variables is given as in Eq.~\eqref{eq:defMarginalPvh}: $p(\bs{v},\bs{h}) = \exp(-E(\bs{v},\bs{h}))/Z$.

This proposed RBM is termed the \textit{gamma-Bernoulli RBM} because its conditional probabilities are given by
\begin{align}
    p(\bs{v}|\bs{h}) &= \mathcal{G}(\bs{v}; \mathbf{V} \bs{h}, -\mathbf{W} \exp(\bs{h})),
    \label{eq:p(v|h)proposed(gamma)}\\
    p(\bs{h}|\bs{v}) &= \mathcal{B}(\bs{h}; \sigmoid[\, (e-1)(\bs{c}+\mathbf{W}^\mathsf{T} \bs{v}) \nonumber\\
    &\hspace{104pt}+\bs{d} +  \mathbf{V}^\mathsf{T} \log(\bs{v})\,]),
    \label{eq:p(h|v)proposed}
\end{align}
where $\mathcal{G}(\:\cdot\:; \bs{\alpha}, \bs{\beta})$ is the element-wise i.i.d.~gamma distribution with a shape-parameter vector $\bs{\alpha}\in\mathbb{R}^I_{++}$ and a rate-parameter vector $\bs{\beta}\in\mathbb{R}_{++}^I$, i.e., $\mathcal{G}(\bs{x}; \bs{\alpha}, \bs{\beta}) = \prod_d \mathcal{G}_d(x_d; \alpha_d, \beta_d)$ with
\begin{equation}
    \mathcal{G}_d(x; \alpha, \beta) = \frac{\beta^\alpha}{\Gamma(\alpha)} x^{\alpha-1} e^{-\beta x},
\end{equation}
and $\Gamma(\cdot)$ is the gamma function.

The gamma distribution is a natural choice for modeling positive data.
Furthermore, some research has reported that the gamma distribution can approximate the distribution of speech signals better than the Gaussian distribution regardless of the type of speech parameterization \cite{gazor2003speech,shin2005statistical,martin2002speech,andrianakis2006mmse}.
Thus, the proposed gamma-Bernoulli RBM should be more suitable for modeling amplitude spectra than the Gaussian-Bernoulli RBM.

\subsection{Implementation of Gamma-Bernoulli RBM}
In the proposed formulation, $\mathbf{V}\bs{h}$ and $-\mathbf{W}\exp(\bs{h})$ correspond to the parameters of the gamma distribution ($\bs{\alpha}$ and $\bs{\beta}$, respectively) as in Eq.~\eqref{eq:p(v|h)proposed(gamma)}.
Therefore, both vectors must be positive for satisfying the definition of the gamma distribution.
To ensure positivity of $\mathbf{V}$ and $-\mathbf{W}$ without causing instability of the training, we parameterize them as follows \cite{cho2011improved}:
\begin{equation}
    \mathbf{W} = -\exp(\widetilde{\mathbf{W}}), \qquad\quad
    \mathbf{V} = \exp(\widetilde{\mathbf{V}}),
\end{equation}
where $\widetilde{\mathbf{W}}\in\mathbb{R}^{I\times J}$, and  $\widetilde{\mathbf{V}}\in\mathbb{R}^{I\times J}$.

Moreover, in order to avoid $\mathbf{V}\bs{h}=\bs{0}$ which occurs when $\bs{h}=\bs{0}$, one may modify the vector $\bs{s}$ given in Eq.~\eqref{eq:defSs} as $\bs{s}=[\,(-\bs{1}+\varepsilon)^\mathsf{T},\bs{d}^\mathsf{T}\,]^\mathsf{T}$ with a small constant $\varepsilon>0$.
This addition makes the shape parameter of the gamma distribution in Eq.~\eqref{eq:p(v|h)proposed(gamma)}, $\bs{\alpha}=\mathbf{V}\bs{v}+\varepsilon$,  always positive as required by the definition.
However, such modification is not so important for practical applications because $\bs{h}=\bs{0}$ rarely happens.

\subsection{Objective Function and Parameter Optimization}
Like the conventional Boltzmann machines, the objective of the proposed RBM is to maximize the log-likelihood:
\begin{align}
    L(\{\bs{v}^{(n)}\}_n) &= \frac{1}{N} \sum_n \log( p(\bs{v}^{(n)})) \\
    &= \frac{1}{N} \sum_n \log\Bigl( \sum_{\bs{h}^{(n)}\!\!\!\!} p(\bs{v}^{(n)},\bs{h}^{(n)})\Bigr) \\
    &= \frac{1}{N} \sum_n \log\Bigl( \sum_{\bs{h}^{(n)}\!\!\!\!} e^{-E(\bs{v}^{(n)},\bs{h}^{(n)})}\Bigr) - \log Z,
\end{align}
where $\bs{v}^{(n)}$ and $\bs{h}^{(n)}$ are the $n$th training data and the corresponding hidden variables, respectively, and $\sum_{\bs{h}^{(n)}}$ represents marginalization over all possible states of $\bs{h}^{(n)}$.

For the optimization, the gradient of the log-likelihood function w.r.t.~the parameters $\bs{\theta}=(\widetilde{\mathbf{W}},\widetilde{\mathbf{V}},\bs{c},\bs{d})$ is required.
Although it can be explicitly written as
\begin{equation}
    \frac{\partial L}{\partial \bs{\theta}} = \Bigl\langle - \frac{\partial E}{\partial\bs{\theta}} \Bigr\rangle_{\!\mathrm{data}} \!-\; \Bigl\langle - \frac{\partial E}{\partial \bs{\theta}} \Bigr\rangle_{\!\mathrm{model}},
\end{equation}
this gradient is practically intractable owing to the second term, where $\langle\cdot\rangle_\mathrm{data}$ and $\langle\cdot\rangle_\mathrm{model}$ represent the expectations on data and model distributions, respectively.
Therefore, as usual in the conventional Boltzmann machines, the contrastive divergence method \cite{hinton2006fast} is applied to approximate the gradient:
\begin{equation}
    \frac{\partial L}{\partial \bs{\theta}} = \Bigl\langle - \frac{\partial E}{\partial\bs{\theta}} \Bigr\rangle_{\!\mathrm{data}} \!-\; \Bigl\langle - \frac{\partial E}{\partial \bs{\theta}} \Bigr\rangle_{\!\mathrm{recon}},
\end{equation}
where $\langle\cdot\rangle_\mathrm{recon}$ is the expectation on the reconstructed data usually obtained through the Gibbs sampling.

The negative partial gradients of the energy function in Eq.~\eqref{eq:defGammaRBMenergy} w.r.t. each parameter are obtained as follows:
\begin{align}
    -\frac{\partial E_{\Gamma\!}}{\partial\widetilde{\mathbf{W}}} &= \mathbf{W} \circ (\bs{v} \exp(\bs{h})^\mathsf{T}), 
    &-\frac{\partial E_{\Gamma\!}}{\partial \bs{c}} &= \exp(\bs{h}),\\
    -\frac{\partial E_{\Gamma\!}}{\partial \widetilde{\mathbf{V}}} &= \mathbf{V} \circ (\log(\bs{v})\, \bs{h}^\mathsf{T}), 
    &-\frac{\partial E_{\Gamma\!}}{\partial \bs{d}} &= \bs{h},
\end{align}
where $\circ$ denotes the element-wise multiplication.

\section{Experiment}
\label{sec:experiments}
The effectiveness of the proposed model was investigated by a speech representation experiment as follows.

\subsection{Configuration}
In the experiment, the ATR speech corpus (set \texttt{B}, speaker \texttt{FTK}) was utilized.
The speech signals of 50 sentences (\texttt{SDA}) were utilized for training, while the other 53 sentences (\texttt{SDJ}) were used for evaluation.
Those signals originally sampled at 20 kHz were downsampled to 16 kHz for speeding up the computation.
The short-time Fourier transform (STFT) was implemented with a 256-sample-long Hamming window and a hop size of 64 samples.
The 129-dimensional data vector $\bs{v}^{(n)}$ was calculated by taking the absolute value of the spectrum of each windowed segment.
After discarding silent segments, the number of the data samples for training was 51\,197.

The proposed RBM was compared with the Gaussian-Bernoulli RBM.
The training data were normalized so that the data distribution was standardized for each RBM.
For the ordinary Gaussian-Bernoulli RBM, as usual, each dimension was normalized so that the data were distributed with center 0 and standard deviation 1.
For the proposed gamma-Bernoulli RBM, each dimension was normalized as
\begin{equation}
    \widetilde{x} = \hat{\beta} x = \frac{\alpha x}{\overline{x}},
\end{equation}
so that the gamma distribution, assumed as $\mathcal{G}(x;\alpha,\hat{\beta})$, becomes the standard form, $\mathcal{G}(\widetilde{x};\alpha,1)$, where $\overline{x}$ denotes the mean of $x$, and $\hat{\beta}$ is the maximum-likelihood estimation of $\beta$.
In this experiment, we considered $\alpha=1$ for the normalization.

Both RBMs were trained by the Adam optimizer \cite{kingma2015adam} with a batch size 100 and a learning rate 0.01.
The number of hidden units was set to 100, 200, 400, or 800.
After training with 100 epochs, the amplitude spectrogram of the evaluation data were encoded and reconstructed using the trained models by calculating the expectation of $\bs{v}$ from the expectation of the encoded signal $\bs{h}$ obtained from the inputted data samples, i.e., reconstruction is obtained from $p(\bs{v}|\mathbb{E}_{ p(\bs{h}|\bs{v}^{(n)})}[\bs{h}])$.
Their performances were evaluated by PESQ and MSE after canceling the effect of normalization by the inverse operation.

\subsection{Results}
We show the experimental results in the following three ways: scores of PESQ, an example of a reconstructed spectrogram, and learning curves in terms of MSE.

\begin{figure}[t!]
  \centering
  \includegraphics[width=0.85\columnwidth]{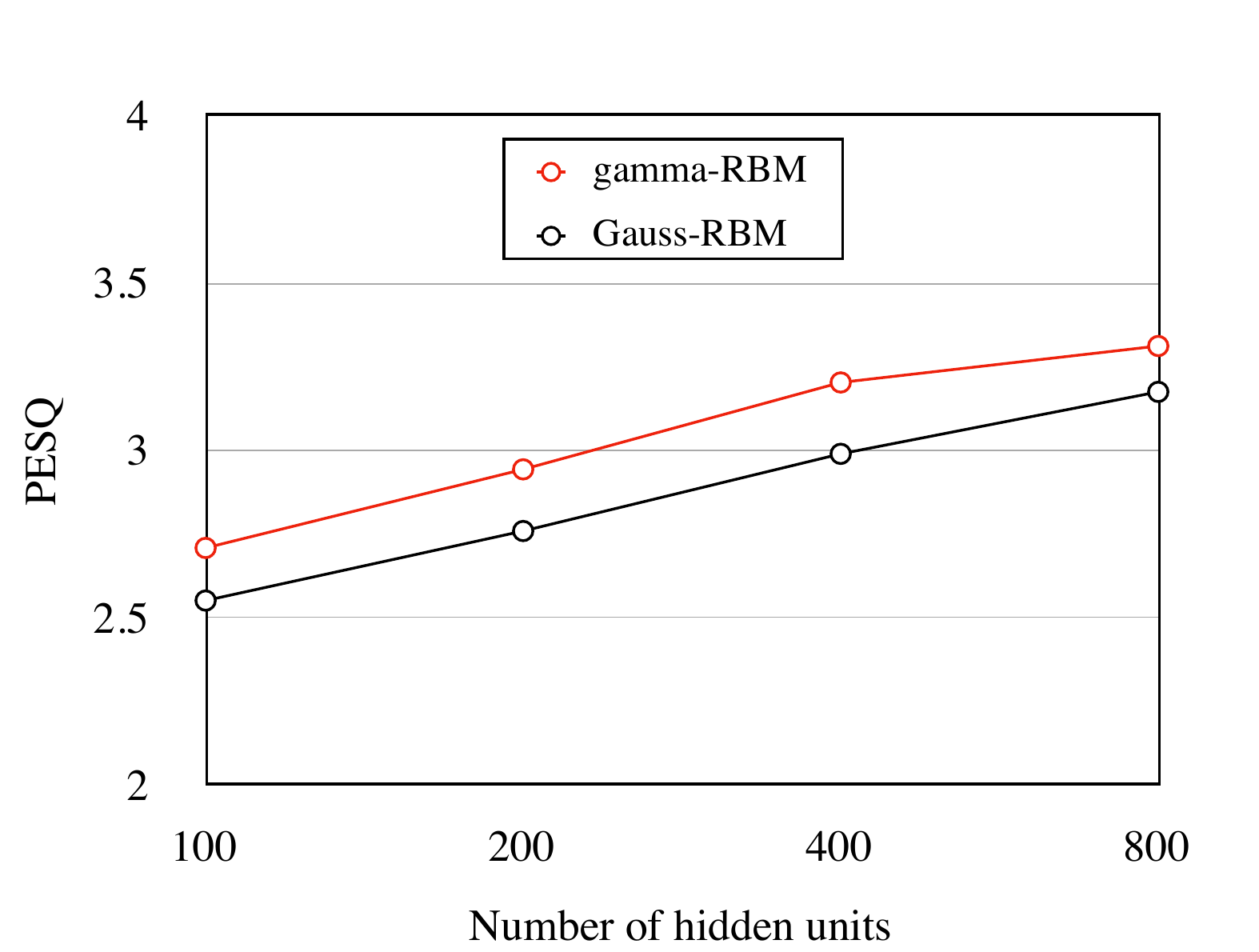}
 \vspace{-4pt}
  \caption{PESQ scores obtained by the proposed (\texttt{gamma-RBM}, red) and conventional (\texttt{Gauss-RBM}, black) models constructed and trained with various numbers of hidden units.}
  \label{fig:pesq}
\end{figure}

Firstly, PESQ scores averaged over all evaluation data are shown in Fig.\:\ref{fig:pesq}.
After reconstructing the amplitude spectrograms, the corresponding signals in the time domain were calculated by the inverse STFT using phase of the original signals.
Then, the PESQ scores were calculated using the original signals as the references.
As illustrated in the figure, the proposed RBM (\texttt{gamma-RBM}) outperformed the ordinary RBM (\texttt{Gauss-RBM}) for all situations.
This should be because the proposed model explicitly considers the log-amplitude spectrogram which is more relevant to the human auditory system.
The proposed model could obtain better scores by increasing the number of hidden units as it did not reach the ceiling with 800 units.

\begin{figure*}[t!]
  \includegraphics[width=1.97\columnwidth]{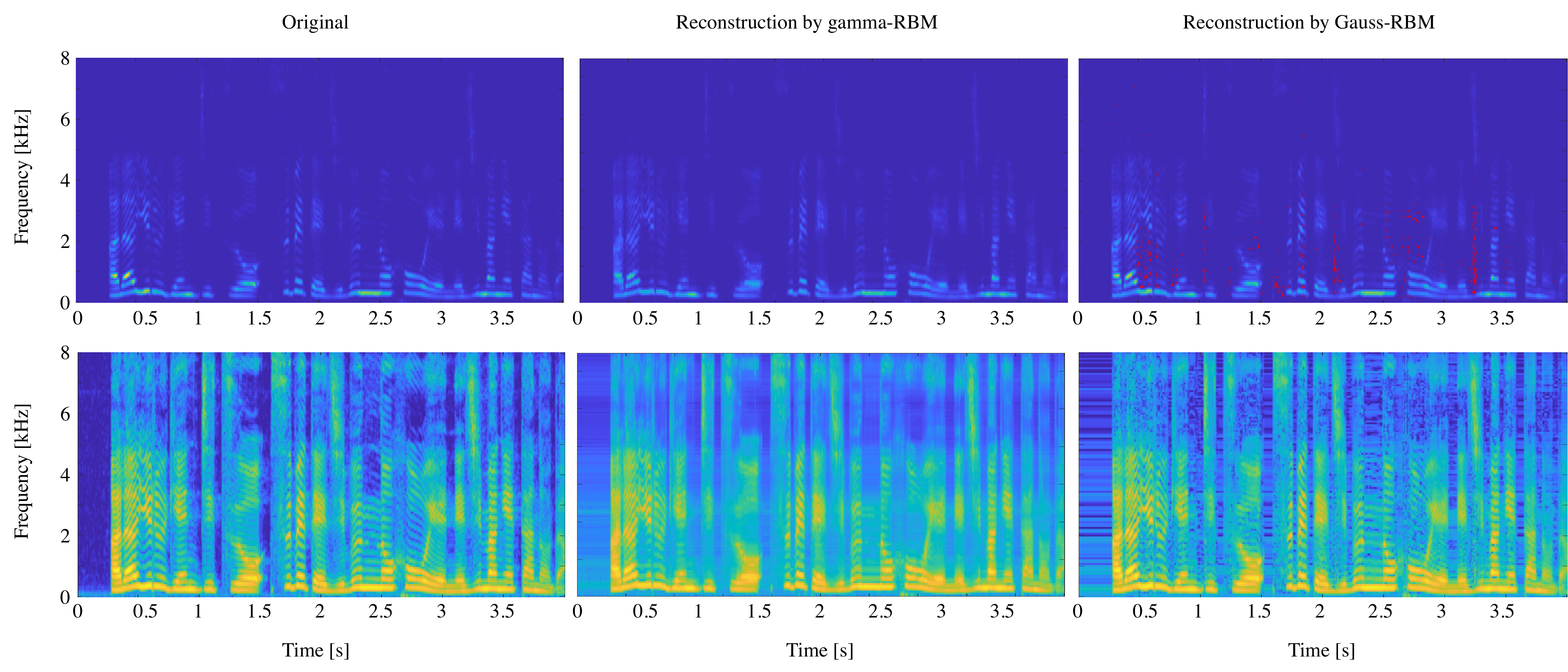}
 \vspace{-7pt}
  \caption{Top and bottom rows show linear- and log-amplitude spectra. From left to right: original, reconstructions by the proposed (\texttt{gamma-RBM (H800)}) and by the conventional (\texttt{Gauss-RBM (H800)}) models. The red regions in the top row represent negative values which should not exist.}
 \vspace{-10pt}
  \label{fig:recon}
\end{figure*}

Secondly, an example of the reconstructed amplitude spectrograms is shown in Fig.\:\ref{fig:recon}, where the number of hidden units was set to 800 (\texttt{H800}).
As can be seen from the top right figure, the conventional model resulted in the negative values indicated by the red points.
Some reconstruction error at the time-frequency bins having small energy can also be noticed in the bottom right figure.
In contrast, the proposed model did not produce any negative value as expected by the definition.
Although the reconstructed spectrogram was smother as shown in the central figure, its spectral envelope seems to be closer to the original signal than the conventional method, which should be the reason of the better PESQ scores.

\begin{figure}[t]
  \centering
  \includegraphics[height=0.6\columnwidth]{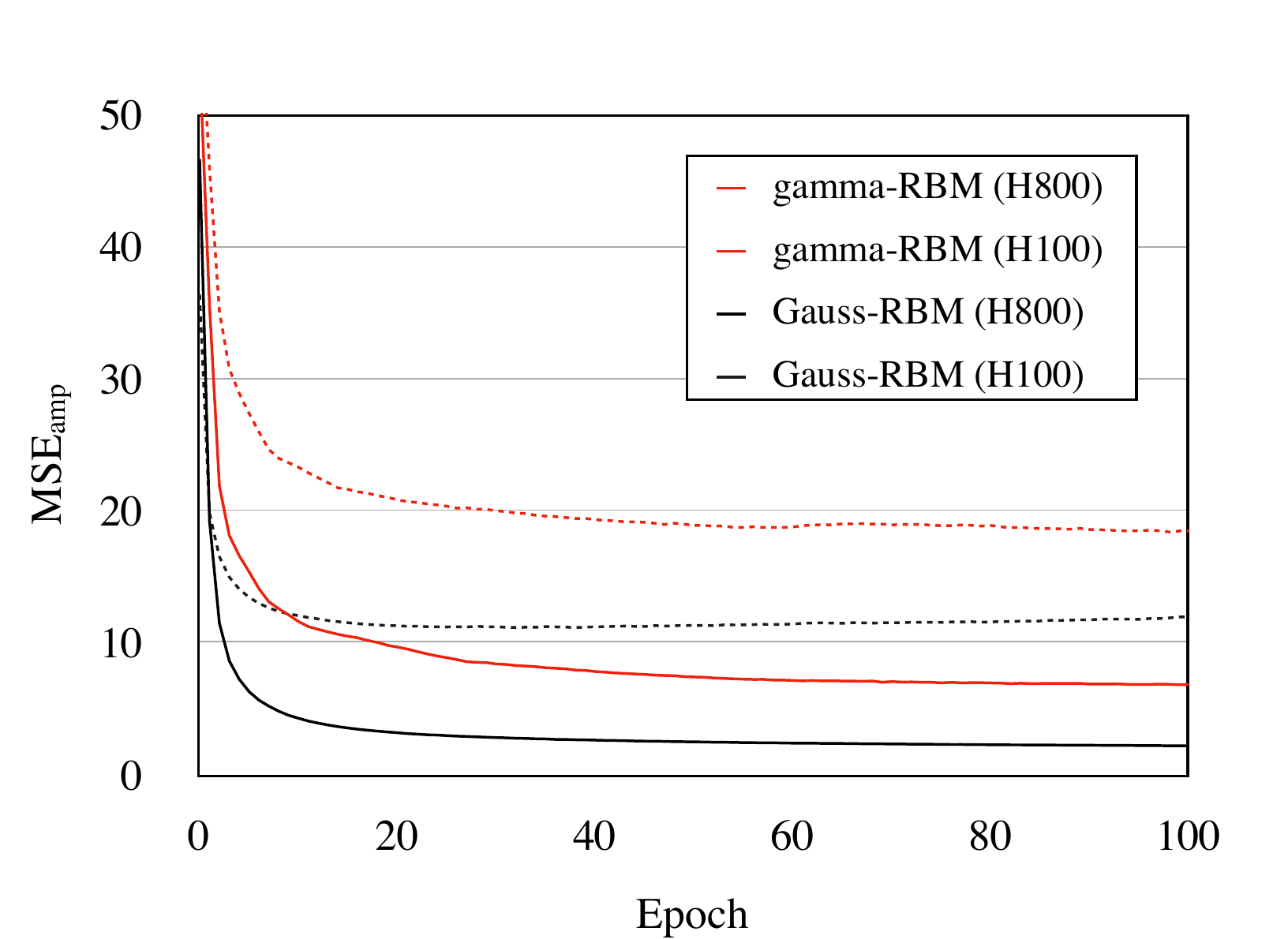}
 \vspace{-4pt}
  \caption{MSE curves w.r.t. the amplitude spectrograms during training of the proposed (\texttt{gamma-RBM}, red) and the conventional (\texttt{Gauss-RBM}, black) models. The numbers after \texttt{H} indicate the number of hidden units.}
  \label{fig:mse_amp}
\end{figure}

\begin{figure}[t]
  \centering
  \includegraphics[height=0.6\columnwidth]{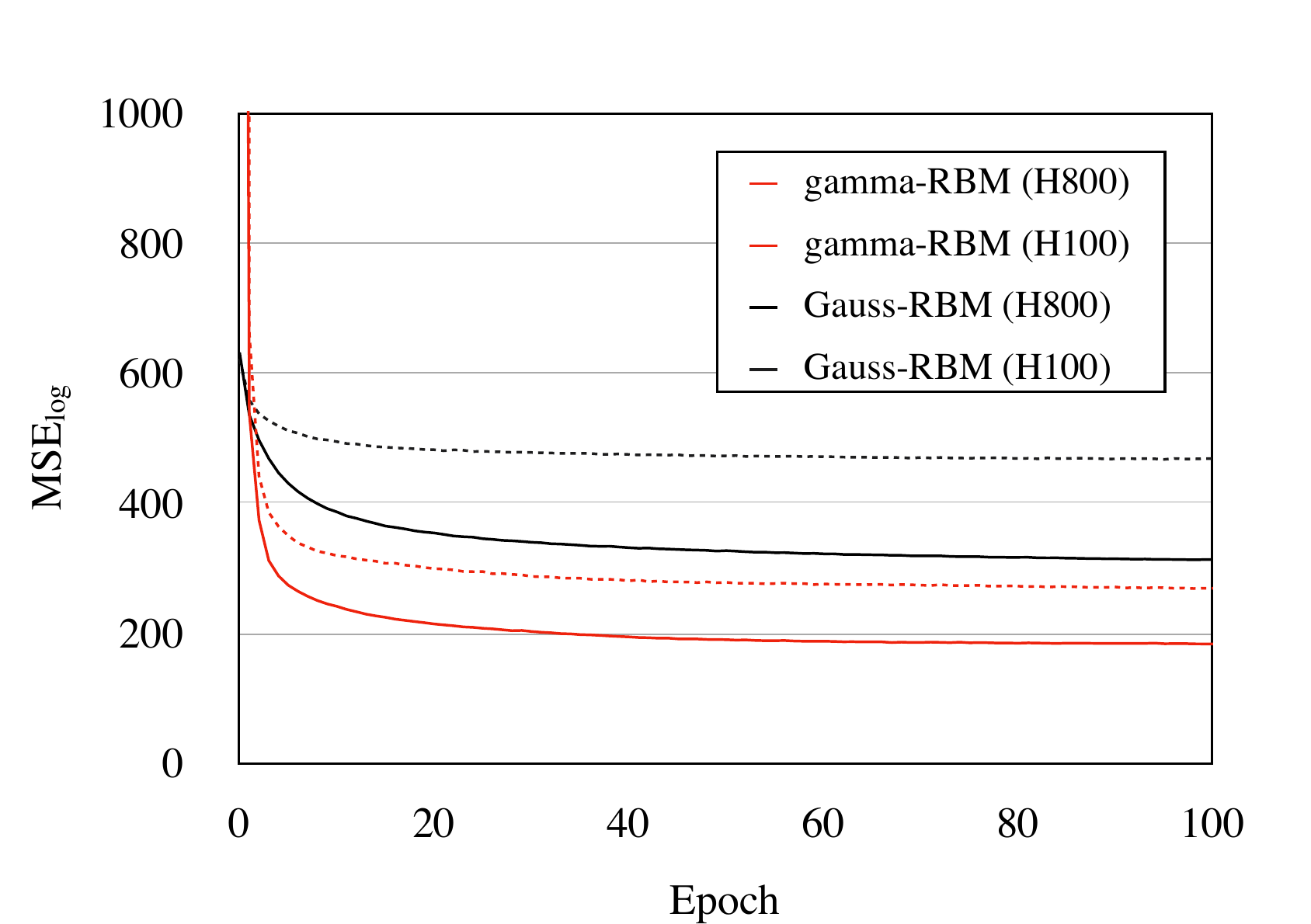}
 \vspace{-4pt}
  \caption{MSE curves w.r.t. the log-amplitude spectrograms during training of the proposed (\texttt{gamma-RBM}, red) and the conventional (\texttt{Gauss-RBM}, black) models. The numbers after \texttt{H} indicate the number of hidden units.}
  \label{fig:mse_log}
\end{figure}

Finally, MSE w.r.t. linear- and log-amplitude spectrograms per epoch are illustrated in Figs.\:\ref{fig:mse_amp} and \ref{fig:mse_log}, respectively.
Since the proposed RBM considers both linear- and log-amplitude spectra, MSE was calculated in both domains as follows:
\begin{align}
{\rm MSE_{amp}} &= \frac{1}{N} \sum_{n=1}^N \bigl\|\, \bs{v}^{(n)} - \hat{\bs{v}}^{(n)} \bigr\|_2^2, \\
{\rm MSE_{log}} &= \frac{1}{N} \sum_{n=1}^N \bigl\|\, \log |\bs{v}^{(n)}| - \log |\hat{\bs{v}}^{(n)}| \,\bigr\|_2^2,
\end{align}
where $\bs{v}^{(n)}$ and $\hat{\bs{v}}^{(n)}$ denote the $n$th original and reconstructed amplitude spectra, respectively, and $N$ is the total number of the segments.
While the conventional models (black) were slightly better than the proposed models (red) in terms of $\mathrm{MSE_{amp}}$ (Fig.\:\ref{fig:mse_amp}), the proposed models outperformed the conventional models in terms of $\mathrm{MSE_{log}}$ (Fig.\:\ref{fig:mse_log}).
By paying attention to the number of hidden units, in Fig.\:\ref{fig:mse_amp}, the proposed model with 800 units (solid red) easily outperformed the conventional model with 100 units (dotted black).
In contrast, in Fig.\:\ref{fig:mse_log}, the conventional model with 800 units (solid black) could not outperform the proposed model with 100 units (dotted red).
These results indicate that the simultaneous consideration of linear- and log-amplitude spectra in the proposed gamma-Bernoulli RBM can improve the overall performance.

\section{Conclusions}
In this paper, we proposed a novel RBM named gamma-Bernoulli RBM.
By modeling data via the gamma distribution, the proposed RBM can naturally handle positive data such as amplitude spectra.
Since it optimizes the parameters by simultaneously considering data in the linear and logarithmic scales, the obtained model should be suitable for an application sensitive to logarithmic quantities as well as the linear scale.

\section*{Acknowledgment}
This work was partially supported by JSPS KAKENHI Grant Number 18K18069.

\bibliographystyle{IEEEtran}
\bibliography{apsipa2020}

\end{document}